\documentclass[superscriptaddress,showpacs,amsmath,amssymb,twocolumn]{revtex4-1}
\usepackage{amsmath, amsthm, amssymb, braket}

\usepackage{amsfonts}

\usepackage{graphicx}
\usepackage{dcolumn}
\usepackage{bm}

\usepackage{color}

\usepackage{ulem}

\begin{document}

\title{Enhanced collectivity of $\gamma$ vibration in neutron-rich Dy isotopes with $N=108 - 110$
}

\author{Kenichi Yoshida}
\affiliation{Graduate School of Science and Technology, Niigata University, Niigata 950-0913, Japan}
\affiliation{Center for Computational Sciences, University of Tsukuba, Tsukuba 305-8577, Japan
}
\author{Hiroshi Watanabe}
\affiliation{School of Physics and Nuclear Energy Engineering, Beihang University, Beijing 100191, China}
\affiliation{RIKEN Nishina Center for Accelerator-Based Science, Wako, Saitama  351-0198, Japan
}%

\date{\today}

\begin{abstract}
{\noindent {\bf Background:} 
The $\gamma$ vibrational mode of excitation is an acknowledged collective mode in deformed nuclei. 
The collectivity depends on the details of the shell structure around the Fermi levels, 
in particular the presence of the orbitals that have the enhanced transition matrix elements of the non-axial quadrupole excitation. 
Quite recently, a sudden decrease in the excitation energy of the $\gamma$ vibration was observed at RIKEN RIBF 
for the neutron-rich Dy isotopes at $N=106$. \\
{\bf Purpose:} 
In the present work, by studying systematically 
the microscopic structure of the $\gamma$ vibration in the neutron-rich Dy isotopes 
with $N=98-114$, we try to understand the mechanism of the observed softening.\\
{\bf Methods:} 
The low-frequency modes of excitation in the neutron-rich rare-earth nuclei are described based on 
nuclear density-functional theory. 
We employ the Skyrme energy-density functionals (EDF) in the Hartree-Fock-Bogoliubov calculation 
for the ground states and in the Quasiparticle Random-Phase Approximation (QRPA) for the excitations.\\
{\bf Results:} 
The lowering of the excitation energy around $N=106$ 
is reproduced well by employing the SkM* and SLy4 functionals. 
It is found that the coherent contribution of the $\nu[512]3/2 \otimes \nu[510]1/2, \nu [510]1/2 \otimes \nu [512]5/2$, 
and $\nu[512]3/2 \otimes \nu[514]7/2$ excitations satisfying the selection rule of the non-axial quadrupole matrix element 
plays a major role in generating the collectivity. We find the similar isotopic dependence of the excitation energy 
in the neutron-rich Er and Yb isotopes as well.\\
{\bf Conclusions:} 
The microscopic framework of the Skyrme-EDF based QRPA describes well 
the isotopic dependence of the energy of the $\gamma$ vibration in the well-deformed neutron-rich rare-earth nuclei. 
The strong collectivity at $N=108-110$ is expected 
as the Fermi level of neutrons lies just among the orbitals that play an important role in generating the collectivity around $N=106$.
}
\end{abstract}

\maketitle

\section{Introduction}\label{intro}

Atomic nuclei reveal spontaneous breaking of the rotational symmetry in both real space and gauge space 
in stepping away from the magic numbers. 
Most of the deformed nuclei still keep the axial symmetry. 
In axially deformed nuclei, a low-frequency quadrupole mode of excitation, 
known as the $\gamma$ vibration~\cite{BM2,RS}, emerges. 
The $\gamma$ vibrational mode of excitation is regarded as a precursory soft mode of 
the permanent non-axial deformation~\cite{BM2}. 

In actual nuclei, the onset of $\gamma$ instability is expected to occur in the transitional regions 
where the nucleus is weakly deformed~\cite{abe90}. 
For example, a schematic Hamiltonian with the pairing-plus-quadrupole residual interactions predicted 
an occurrence of the triaxial deformation with an axial-deformation parameter $\beta < 0.2$ 
in the Os ($Z=76$) and Pt ($Z=78$) isotopes around $A=188$, and 
the lowering of the second $2^+$ ($2^+_2$) state in the region of $A=186-190$~\cite{kum68}.

Regardless of the nature of axial asymmetry, being either $\gamma$ vibration, $\gamma$ softness (instability), 
or permanent triaxial deformation, 
in deformed even-even nuclei the energy of the $2^+_2$ state, 
on which a characteristic $K^{\pi} = 2^+$ $\gamma$ band is built, 
decreases as the non-axial collectivity is enhanced~\cite{sat10}. 
Experimentally, the $2^+_2$ states were observed at low excitation energy ($\lesssim$ 800 keV) 
for the Dy ($Z=66$) and Er ($Z=68$) isotopes with $N=98$~\cite{nndc}. 
In such well-deformed, axially-symmetric nuclei, the energies of the low-spin members of the $\gamma$-vibrational band 
obey reasonably well the regularity that is expected from the standard rotor picture.
On the other hand, 
some nuclei tend to represent a staggering in energies of the even- and odd-spin levels within a $\gamma$ band. 
The pattern of the even-odd staggering is expected to be opposite for $\gamma$-unstable and $\gamma$-rigid 
rotors~\cite{Wil56,Dav58}. 
For Pt isotopes, the property of the $\gamma$ band in $^{184}$Pt ($N=106$) 
is close to the $\gamma$-unstable case, while the other isotopes, in particular around $N=116$, 
exhibit behaviors that fall between the two extremes~\cite{nndc}. 
Although it is not simple to distinguish the $\gamma$ softness and the rigid triaxial deformation 
in terms of the staggering in $\gamma$-band energies even for Os isotopes, 
which are supposed to be the typical $\gamma$-soft nuclei~\cite{Wu96}, 
the fact that the $2^+_2$ state is lower in energy than the $4^+_1$ level of the ground-state rotational 
band in $^{192}$Os ($N=116$) is indicative of strongly enhanced axial asymmetry. 
The $N=116$ isotones, $^{192}$Os and $^{190}$W ($Z=74$), 
the latter having the lowest $2^+_2$ energy (454 keV) in this region~\cite{alk09}, 
reveal isomeric-decay transitions with anomalously low $K$ hindrances~\cite{dra13, Lan10}, 
indicating that $K$ is no longer a good quantum number due to substantial axial asymmetry.

Neutron-rich Dy isotopes around double midshell were produced by in-flight fission of a $^{238}$U beam at the RI-Beam Factory (RIBF) at RIKEN, 
and their level structures were studied using the EURICA setup. Isomers with $K^{\pi} = 6^+$ and $8^-$ 
were identified in $^{170}$Dy ($N=104$)~\cite{Sod16} and $^{172}$Dy ($N=106$)~\cite{wat16}, respectively, 
both of which decay to the respective ground-state and $\gamma$ bands. 
Despite the robust nature of the ground-state rotational band and the $K$ isomer, 
being characteristic of an axially deformed nucleus, $^{172}$Dy has the $\gamma$ band at unusually low excitation energy 
compared to the nearby well-deformed nuclei~\cite{wat16}. 
This spectroscopic result suggests that the shell structure near the Fermi surface plays crucial roles in developing the collectivity 
in this doubly mid-shell region. 

The vibrational mode of excitation is described microscopically 
by the Random Phase Approximation (RPA) 
on top of the mean-field ground state~\cite{RS}.
Since the ground state is deformed and in the superfluid phase, 
the deformed Quasiparticle-RPA (QRPA) is commonly employed to study the $\gamma$ vibration~\cite{bes65}. 
The $\gamma$ vibration in the rare-earth nuclei 
was investigated systematically in a Skyrme energy-density functional (EDF) approach 
and comparison with the experimental data was made in Refs.~\cite{ter11,nes16}.
The authors of Refs.~\cite{ter11,nes16} found that the SkM* functional~\cite{bar82} 
reproduces well the trends in energies and transition probabilities in the light rare-earth nuclei, 
although the numerical results obtained by the QRPA are not perfect.

In this article, we investigate systematically 
the microscopic structure of the $\gamma$ vibration in the neutron-rich Dy isotopes 
with $N=98-114$ in a Skyrme EDF approach. 
The $\gamma$ vibration is described by the deformed HFB + QRPA. 
This work is thus an extension of the study made in Ref.~\cite{ter11} to the neutron rich side. 
We discuss the microscopic mechanism of the decrease in the excitation 
energy at $N=106$ observed experimentally. 

In Sec.~\ref{method}, we introduce the basic equations of the deformed HFB + QRPA 
needed to investigate the microscopic structure of the $\gamma$ vibration. 
The results are presented and discussed in Sec.~\ref{result}. 
Finally, we summarize this article in Sec.~\ref{summary}. 

\section{Numerical method}\label{method}

\subsection{Basic equations of deformed HFB + QRPA}
Details of the axially deformed HFB in the cylindrical-coordinate space
with the Skyrme EDF and the QRPA in the quasiparticle (qp) representation
can be found in Refs.~\cite{yos08,yos13}. 
Here, we briefly recapitulate the outline of the formulation.

To describe the nuclear deformation 
and the pairing correlations simultaneously, taking into account the spatial extension,
we solve the HFB equations~\cite{dob84,bul80}
\begin{align}
\begin{pmatrix}
h^{q}(\boldsymbol{r}\sigma)-\lambda^{q} & \tilde{h}^{q}(\boldsymbol{r}\sigma) \\
\tilde{h}^{q}(\boldsymbol{r}\sigma) & -[h^{q}(\boldsymbol{r}\sigma)-\lambda^{q}]
\end{pmatrix}
\begin{pmatrix}
\varphi^{q}_{1,\alpha}(\boldsymbol{r}\sigma) \\
\varphi^{q}_{2,\alpha}(\boldsymbol{r}\sigma)
\end{pmatrix} \notag \\
= E_{\alpha}
\begin{pmatrix}
\varphi^{q}_{1,\alpha}(\boldsymbol{r}\sigma) \\
\varphi^{q}_{2,\alpha}(\boldsymbol{r}\sigma)
\end{pmatrix} \label{HFB_equation}
\end{align}
in real space using cylindrical coordinates $\boldsymbol{r}=(\rho,z,\phi)$. 
Here, $q=\nu$ (neutron) or $\pi$ (proton). 
We assume axial and reflection symmetries.
Since we consider the even-even nuclei only, 
the time-reversal symmetry is also assumed. 
A nucleon creation operator $\hat{\psi}^{\dagger}(\boldsymbol{r}\sigma)$ 
at the position $\boldsymbol{r}$ with the intrinsic spin $\sigma$ is  
written in terms of the qp wave functions as
\begin{equation}
\hat{\psi}^{\dagger}_q(\boldsymbol{r}\sigma)
=\sum_{\alpha}\varphi^q_{1,\alpha}(\boldsymbol{r}\bar{\sigma})\hat{\beta}^{\dagger}_{q,\alpha}
+\varphi_{2,\alpha}^{q*}(\boldsymbol{r}\sigma)\hat{\beta}_{q,\alpha}
\end{equation}
with the quasiparticle creation and annihilation operators $\hat{\beta}^\dagger, \hat{\beta}$.
The notation $\varphi(\boldsymbol{r}\bar{\sigma})$ is defined by 
$\varphi(\boldsymbol{r}\bar{\sigma})=-2\sigma \varphi(\boldsymbol{r}-\sigma)$. 

For the mean-field Hamiltonian $h$, we mainly employ the SkM* functional~\cite{bar82}. 
For the pairing energy, we adopt the volume-type pairing following Ref.~\cite{ter11}; 
the strength is fixed as $V_0 = -190$ MeV fm$^3$ and $-220$ MeV fm$^3$ for neutrons and protons, respectively.
Since we assume the axially symmetric potential,
the $z-$component of the qp angular momentum, $\Omega$,
is a good quantum number. 
Assuming time-reversal symmetry and reflection symmetry
with respect to the $x-y$ plane,
the space for the calculation can be reduced into the one with
positive $\Omega$ and positive $z$ only. 

Using the qp basis obtained
as a self-consistent solution of the HFB equations (\ref{HFB_equation}),
we solve the QRPA equation in the matrix formulation~\cite{row70}
\begin{equation}
\sum_{\gamma \delta}
\begin{pmatrix}
A_{\alpha \beta \gamma \delta} & B_{\alpha \beta \gamma \delta} \\
-B_{\alpha \beta \gamma \delta} & -A_{\alpha \beta \gamma \delta}
\end{pmatrix}
\begin{pmatrix}
X_{\gamma \delta}^{i} \\ Y_{\gamma \delta}^{i}
\end{pmatrix}
=\hbar \omega_{i}
\begin{pmatrix}
X_{\alpha \beta}^{i} \\ Y_{\alpha \beta}^{i}
\end{pmatrix} \label{eq:AB1}.
\end{equation}
The residual interaction in the particle-hole (p-h) channel appearing
in the QRPA matrices $A$ and $B$ is
derived from the Skyrme EDF. 
The residual Coulomb interaction is neglected because of the computational limitation.
We expect that the residual Coulomb interaction plays only a minor role~\cite{ter05,sil06,eba10,nak11}, 
in particular on the $K^{\pi}=2^+$ states that is orthogonal to the spurious modes.
We also drop the so-called $``{J}^{2}"$ term $C_{t}^{T}$ both in 
the HFB and QRPA calculations for the selfconsistency. 
The residual interaction in the
particle-particle (p-p) channel is the same one used in the HFB calculation.

The electric and neutron reduced probabilities are evaluated 
with the intrinsic transition strengths to 
the $\gamma$-vibrational mode $i$ in the rotational coupling scheme~\cite{BM2} as 
\begin{align}
B(\mathrm{E}2; 0^{+}_{\mathrm{gs}} \to 2^{+}_\gamma) &=2 e^2 |\langle i|\hat{F}^{\pi}_{\lambda=2, K=2}|0\rangle|^{2}, \\
B(\mathrm{N}2; 0^{+}_{\mathrm{gs}} \to 2^{+}_\gamma) &= 2|\langle i|\hat{F}^{\nu}_{\lambda=2, K=2}|0\rangle|^{2},
\end{align}
where
\begin{equation}
\hat{F}^q_{\lambda=2,K=2} = \sum_{\sigma}\int d\boldsymbol{r} r^{2}Y_{2 2}(\hat{r})
\hat{\psi}_{q}^{\dagger}(\boldsymbol{r}\sigma) \hat{\psi}_{q}(\boldsymbol{r}\sigma).
\label{operator}
\end{equation}
The intrinsic transition matrix element is given as
\begin{align}
\langle i | \hat{F}^q_{\lambda K}|0 \rangle &= \sum_{\alpha \beta} M^{q,i}_{\alpha \beta} \label{ME} \\
& \equiv \sum_{\alpha \beta} (X^i_{\alpha \beta}+Y^i_{\alpha \beta}) \langle \alpha \beta |\hat{F}^q_{\lambda K}|0\rangle
\end{align}
with the RPA amplitudes and the two-quasiparticle (2qp) matrix elements.

\begin{figure}[t]
\begin{center}
\includegraphics[scale=0.28]{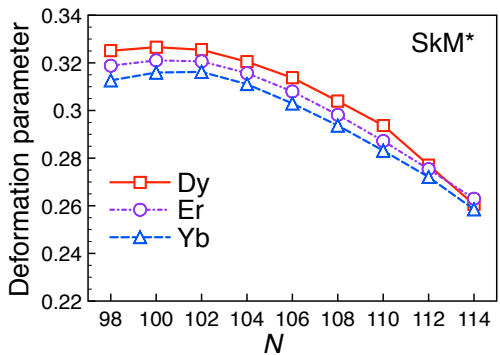}
\caption{(Color online) 
Matter axial-deformation parameters calculated with the SkM* functional for the Dy, Er, and Yb isotopes.}
\label{def}
\end{center}
\end{figure}

\begin{figure*}[t]
\begin{center}
\includegraphics[scale=0.35]{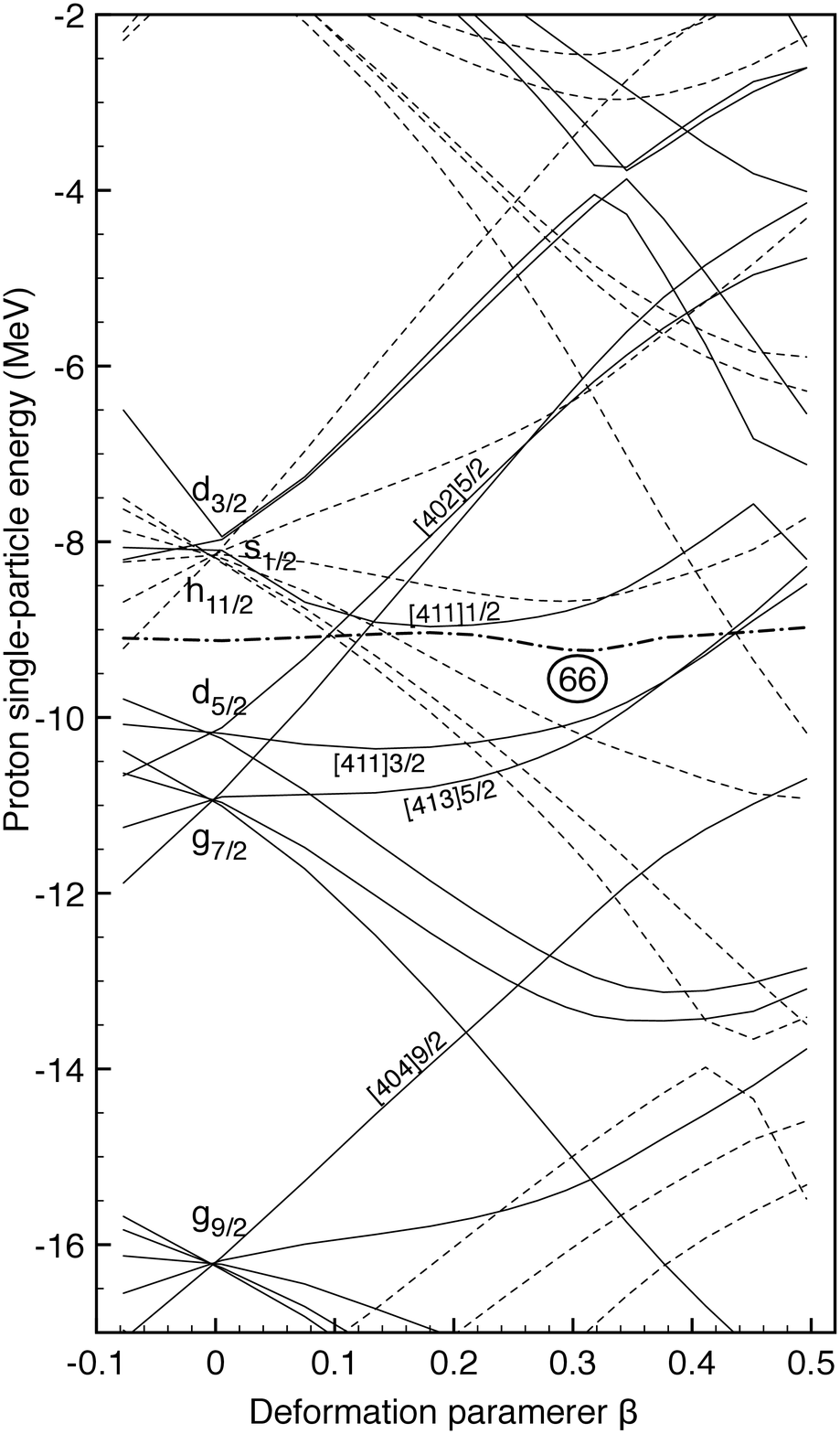}
\includegraphics[scale=0.35]{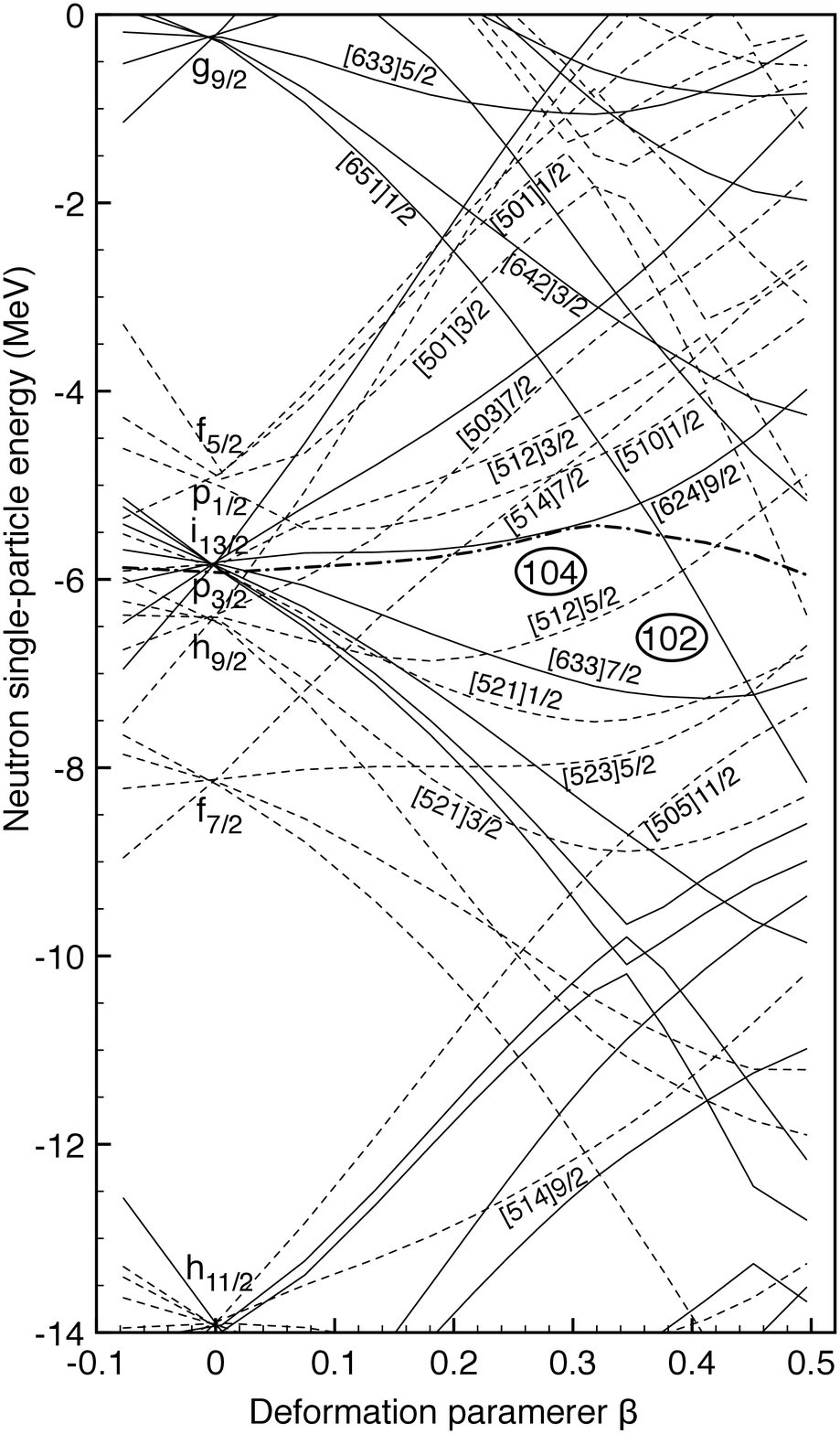}
\caption{Single-particle energies of protons (left) and neutrons (right) as functions of axial-deformation parameter $\beta$ in $^{172}$Dy. 
The positive- and negative-parity states are represented by the solid and dashed lines, and the Fermi levels are denoted 
by the dashed-dotted lines. Some levels relevant to the discussion are indicated by the asymptotic quantum numbers [$N n_3 \Lambda]\Omega$.}
\label{spe}
\end{center}
\end{figure*}

\subsection{Details of the numerical calculation}\label{calculation}

For solution of the HFB equations (\ref{HFB_equation}), 
we use a lattice mesh size $\Delta\rho=\Delta z=0.6$ fm and a box
boundary condition at $\rho_{\mathrm{max}}=14.7$ fm, $z_{\mathrm{max}}=14.4$ fm. 
The differential operators are represented by use of the 13-point formula of finite difference method. 
Since the parity ($\pi$) and  the magnetic quantum number ($\Omega$) 
are good quantum numbers, 
the HFB Hamiltonian 
becomes in a block diagonal form with respect to each $(\Omega^{\pi},q)$ sector.
The HFB equations for each sector are solved independently with 64 cores for 
the qp states up to $\Omega=31/2$ with positive and negative parities. 
Then, the densities and the HFB Hamiltonian are updated, 
which requires communication among the 64 cores.
The modified Broyden's method~\cite{bar08} is utilized to calculate new densities.
The qp states are truncated according to the qp
energy cutoff at $E_\alpha \leq 60$ MeV.

We introduce an additional truncation for the QRPA calculation,
in terms of the 2qp energy as
$E_{\alpha}+E_{\beta} \leq 60$ MeV.
This reduces the number of 2qp states to, for instance,
about 40500 for the $K^{\pi}=2^{+}$ excitation in $^{180}$Dy.
The calculation of the QRPA matrix elements in the qp basis 
is performed in the parallel computers. 
In the present calculation, all the matrix elements are real and 
evaluated by the two-dimensional numerical integration. 
We use 256 cores to compute them.

To save the computing time for diagonalization of the QRPA matrix, 
we employ a technique to reduce the non-Hermitian eigenvalue problem to 
a real symmetric matrix of half the dimension~\cite{ull71,RS}. 
For diagonalization of the matrix, 
we use the ScaLAPACK {\sc pdsyev} subroutine~\cite{scalapack}.
To calculate the QRPA matrix elements and to diagonalize the matrix, 
it takes about 220 core hours and 35 core hours, respectively 
on the COMA(PACS-IX), the supercomputer facility at the CCS in Tsukuba.

\section{Results and discussion}\label{result}

The even-$N$ Dy isotopes with $N=98-114$ under investigation are all well deformed 
with the SkM* functional as shown in Fig.~\ref{def}; 
the matter axial-deformation parameter $\beta$ ranges from 0.26 to 0.33. 
The deformation saturates for $N=98-102$ and decreases as the neutron number increases. 
According to the systematic Skyrme-EDF calculation~\cite{sto03}, 
the deformation gets weaker toward the spherical magic number at $N=126$.
The even-$N$ Er and Yb isotopes with $N=98-114$ are also well deformed 
similarly to the Dy isotopes. 
We assume the nuclei under consideration are described in a strong coupling picture~\cite{BM2}. 
Experimentally, the ratio of the excitation energies of the $4^+_1$ state to the $2^+_1$ state 
stays around 3.3 in the Dy, Er, and Yb isotopes with $N=98-106$~\cite{nndc}.

Figure~\ref{spe} shows the single-particle (sp) energies of protons and neutrons 
as functions of the deformation parameter $\beta$ in $^{172}$Dy. 
The sp energies are obtained by re-diagonalizing the mean-field Hamiltonian $h$ 
in the HFB equation~(\ref{HFB_equation}).
One can see that 
a $Z=66$ deformed-shell gap intervenes 
the $\pi [411]1/2$ and $\pi [411]3/2$ orbitals. 
The $\pi [413]5/2$ orbital is also located around the Fermi level. 
It is noted that the selection rule of the enhanced matrix element for the operator~(\ref{operator})
is given as
\begin{equation}
\Delta N= 0 \hspace{3pt}\mathrm{or}\hspace{3pt} 2, \Delta n_3 =0, \Delta \Lambda = \Delta \Omega= \pm 2.
\label{selection}
\end{equation}
As we will see, the 2qp excitations of $\pi [411]1/2 \otimes \pi [411]3/2$ and 
$\pi [411]1/2 \otimes \pi [413]5/2$ have a significant contribution to generation of the collectivity. 
These 2qp excitation satisfy the selection rule~(\ref{selection}), 
and thus plays a noticeable role in the occurrence of the $\gamma$ vibration in the Dy isotopes under consideration.
Since the shell structure of protons does not alter so much with the neutron number, 
the isotopic dependence of the excitation energy of the $\gamma$ vibration is 
governed by the details of the shell structure of neutrons.

\begin{figure}[t]
\begin{center}
\includegraphics[scale=0.28]{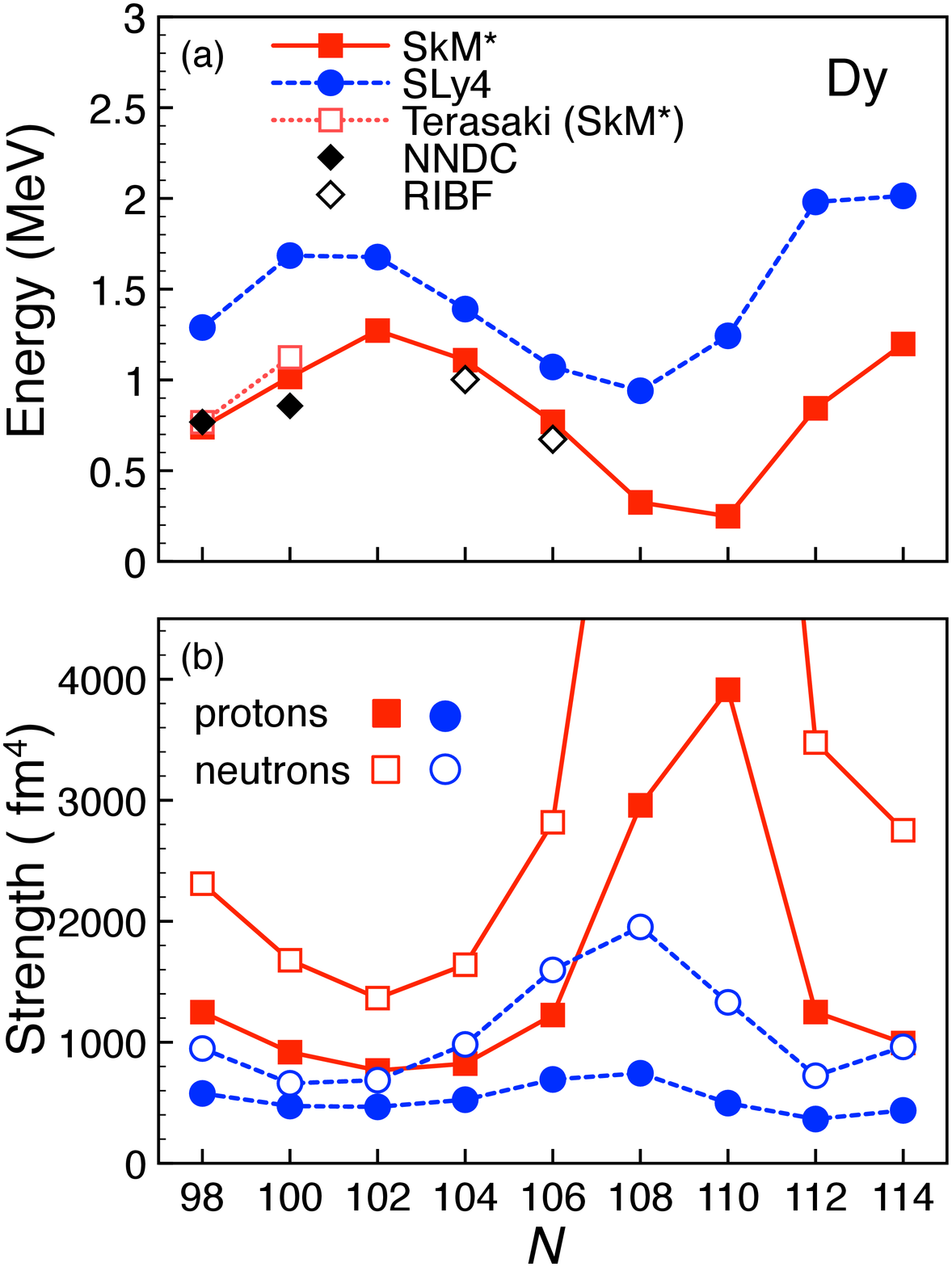}
\caption{(Color online) (a) QRPA frequency of the $\gamma$ vibrational mode obtained 
by using the SkM* and SLy4 functionals. Shown are also the QRPA results in Ref.~\cite{ter11} 
and the experimental data~\cite{nndc, Sod16,wat16}. 
(b) Intrinsic quadrupole transition strengths to the $\gamma$ vibrational mode. 
The electric (proton) and neutron transition strengths are denoted by the filed and open symbols, 
and the results obtained by using the SkM* and SLy4 functionals are denoted by the 
square and circle. 
The neutron transition strength obtained by using the SkM* in $^{174}$Dy and $^{176}$Dy 
is 7670 fm$^4$ and 10830 fm$^4$, respectively.}
\label{Dy_gamma}
\end{center}
\end{figure}

The upper panel in Fig.~\ref{Dy_gamma} shows the QRPA frequency of the $\gamma$ vibration. 
The results obtained by using the SLy4 functional~\cite{cha98} are also shown 
together with 
the numerical results in Ref.~\cite{ter05},  
the experimental data tabulated in Ref.~\cite{nndc}, and the data obtained at RIKEN~\cite{Sod16,wat16}. 
For the data of RIKEN, the band head energy was extrapolated assuming 
the observed $I^{\pi}=5^+, 6^+$, and $7^+$ states belong to the $\gamma$ band. 

We see that both the SkM* and SLy4 functionals reproduce well the lowering of the excitation energy at $N=106$. 
The lower panel in Fig.~\ref{Dy_gamma} shows the intrinsic quadrupole transition strengths to 
the $\gamma$-vibrational mode using the SkM* and SLy4 functionals. 
The electric reduced probabilities calculated with the SkM* and SLy4 functionals 
are about 10 -- 20 in Weisskopf unit (W.u.) 
except for $N=106 - 110$ with the SkM* functional. 
In the case of the SkM* functional, the QRPA frequency drops down to a few hundred keV at $N=108 - 110$, 
and the quadrupole transition strengths increase drastically. 
Indeed, the sum of the backward-going amplitude is 1.90 and 2.59 for $N=108$ and 110, respectively, 
which indicates the RPA overestimates the collectivity of the $\gamma$ vibrational mode. 
Since this kind of singularity occurs artificially due to the small amplitude approximation, 
we are going to discuss the isotopic trend in the energy and transition strength 
rather than putting emphasis on the absolute values.

In $^{164}$Dy, the $\gamma$ vibration is constructed mainly by 
the $\pi [411]1/2 \otimes \pi [411]3/2$ excitation with an amplitude $X^2 - Y^2 = 0.20$, 
and the $\pi [411]1/2 \otimes \pi [413]5/2$ excitation with 0.16. 
In addition to these 2qp excitations of protons, the $\nu [521]1/2 \otimes \nu [523]5/2$ and
$\nu [521]1/2 \otimes \nu [521]3/2$ excitations have 
a predominant contribution with the amplitude of 0.24 and 0.16, respectively. 
These 2qp excitations of neutrons also satisfy the selection rule~(\ref{selection}).

When two neutrons are added to $^{164}$Dy, 
the Fermi level of neutrons in $^{166}$Dy gets higher. 
Then, the 2qp excitation of neutrons, that play a major role in $^{164}$Dy, 
are a hole-hole type excitation, 
and thus the quadrupole matrix element is reduced. 
Therefore, the collectivity of $\gamma$ vibration gets weaker. 

In $^{168}$Dy, the contribution of the above-mentioned 2qp excitations of neutrons 
is less important. 
The amplitude of the $\nu [521]1/2 \otimes \nu [523]5/2$ and
$\nu [521]1/2 \otimes \nu [521]3/2$ excitations is 0.08 and 0.03 only.
Instead, the $\nu [510]1/2 \otimes \nu [512]5/2$ excitation starts to contribute (with an amplitude of 0.13) 
to the $\gamma$ vibration, 
though this is a particle-particle type excitation. 
The collectivity of $\gamma$ vibration depends on how far the $\nu [510]1/2$ and $\nu [512]5/2$ 
orbitals are located from the Fermi level. 

\begin{figure}[t]
\begin{center}
\includegraphics[scale=0.38]{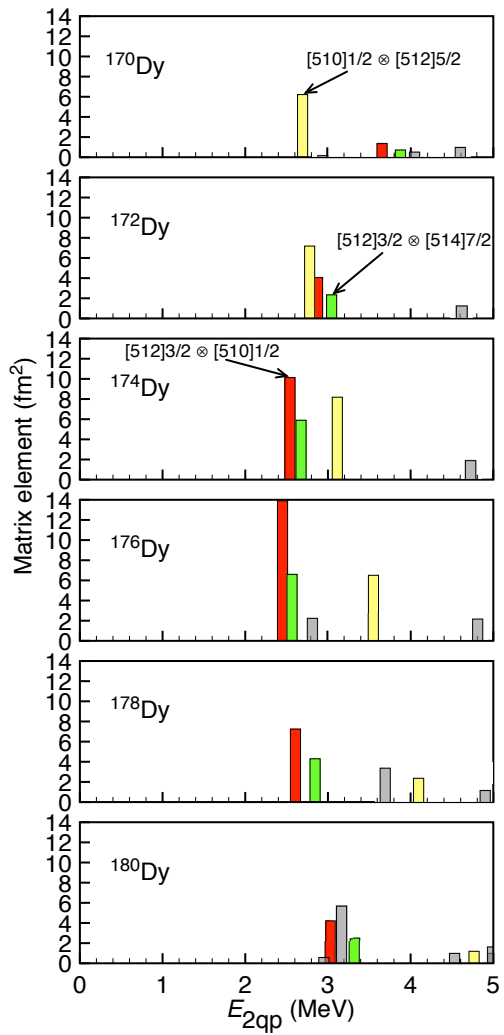}
\caption{(Color online) 
Neutron matrix elements $M^{\nu}_{\alpha \beta}$ of the 2qp excitations near the Fermi level 
for the $\gamma$ vibrational mode in the Dy isotopes.
}
\label{amplitude}
\end{center}
\end{figure}

Beyond $^{168}$Dy, the $\nu [510]1/2 \otimes \nu [512]5/2$ excitation plays a central role 
in generating the $\gamma$ vibration. 
Figure~\ref{amplitude} shows the matrix elements $M_{\alpha \beta}^\nu$ in Eq.~(\ref{ME})
of the 2qp excitations of neutrons near the Fermi level, carrying the low 2qp-excitation energy. 
Actually, in $^{170}$Dy, the $\gamma$ vibration is constructed mainly by 
the $\pi [411]1/2 \otimes \pi [411]3/2$ excitation with an amplitude of 0.28, 
the $\pi [411]1/2 \otimes \pi [413]5/2$ excitation with 0.20, 
and the $\nu [510]1/2 \otimes \nu [512]5/2$ excitation with 0.25. 

When two more neutrons are added to $^{170}$Dy, the particle-particle type excitation, 
such as the $\nu [512]3/2 \otimes \nu [510]1/2$ and $\nu [512]3/2 \otimes \nu [514]7/2$ excitations  
starts to contribute to the $\gamma$ vibration 
coherently as shown in Fig.~\ref{amplitude} with an amplitude of 0.10 and 0.06. 
These 2qp excitations of neutrons again satisfy the selection rule (\ref{selection}). 
This is a microscopic mechanism of the lowering of the excitation energy 
of $\gamma$ vibration in $^{172}$Dy observed recently~\cite{wat16}. 

With an increase in the neutron number, 
the Fermi level of neutrons lies just among 
the $\nu [512]3/2, \nu[510]1/2$, and $\nu[514]7/2$ orbitals in $^{174, 176}$Dy. 
The 2qp matrix element of these excitations thus achieves maximum value in these isotopes. 
Therefore, we obtain the strongest collectivity. 
When more neutrons are added, these 2qp excitations are a hole-hole type excitation, 
and the unperturbed 2qp states are located higher in energy. 
Then, the quadrupole transition strength is smaller, and 
the excitation energy of $\gamma$ vibration is higher. 

\begin{figure}[t]
\begin{center}
\includegraphics[scale=0.29]{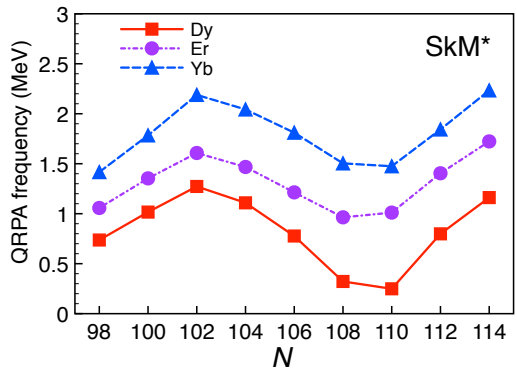}
\caption{(Color online) QRPA frequencies of the $\gamma$ vibrational mode 
 in the Dy, Er, and Yb isotopes with $N=98-114$ obtained by using the SkM* functional.  
}
\label{gamma_energy}
\end{center}
\end{figure}

Finally, we investigate the $\gamma$ vibration in the neutron-rich Er and Yb isotopes. 
Figure~\ref{gamma_energy} shows the QRPA frequency of the $\gamma$ vibration 
in the even-$N$ neutron-rich Er and Yb isotopes together with the Dy isotopes. 
We see that the isotopic trend in the energy is similar to each other. 
Lowering of the energy is seen in $N=98$ and $N=108$,  and 110. 
As we saw in Fig.~\ref{def}, the deformation property of the Er and Yb isotopes is 
similar to that of the Dy isotopes, and the shell structure of neutrons 
near the Fermi level may not be very different from that of the Dy isotopes. 

Indeed, the $\gamma$ vibration in $^{176}$Er ($^{178}$Yb) is coherently constructed by 
the $\nu[512]3/2 \otimes \nu[510]1/2, \nu [512]3/2 \otimes \nu [514]7/2$, and 
$\nu [510]1/2 \otimes \nu[512]5/2$ excitations with a weight of 0.26 (0.42), 0.18 (0.29), and 0.12 (0.11), respectively. 
For the proton configuration, $\pi [411]1/2 \otimes \pi [411]3/2$ and $\pi [411]1/2 \otimes \pi [413]5/2$ 
excitations have some contribution with a weight of 0.17 (0.02) and 0.09 (0.01). 
Since the Fermi level of protons in the Er and Yb isotopes is located higher in energy than the $\pi [411]1/2$ orbital, 
the transition matrix element is reduced, and the protons' contribution to the $\gamma$ vibration is smaller.
Therefore, the collectivity of $\gamma$ vibration is weaker than in the Dy isotopes. 
However, we can say that the microscopic mechanism governing the enhanced collectivity around $N=108$ 
in the Dy isotopes is robust in the neighboring nuclei. 

For the higher-$Z$ nuclei, where the deformation is weak, 
the amplitudes of shape fluctuation about the equilibrium shape increase. 
In such a situation, the large-amplitude motion has to be considered~\cite{mat16}. 

\section{Summary}\label{summary}
We investigated the microscopic structure of the $\gamma$ vibration in the neutron-rich Dy isotopes 
with $N=98-114$ systematically in a Skyrme energy-density functional approach 
to clarify the underling mechanism of the lowering of the excitation energy around $N=106$ observed recently at RIKEN. 
The isotopic dependence of the energy 
is reproduced well by employing the SkM* and SLy4 functionals. 
We found that the coherent contribution of the $\nu[512]3/2 \otimes \nu[510]1/2, \nu [510]1/2 \otimes \nu [512]5/2$, 
and $\nu[512]3/2 \otimes \nu[514]7/2$ excitations satisfying the selection rule of the non-axial quadrupole matrix element 
plays a major role in generating the collectivity. 
Furthermore, we found the similar isotopic dependence of the excitation energy 
in the neutron-rich Er and Yb isotopes as well. 
The strong collectivity at $N=108-110$ is expected 
as the Fermi level of neutrons lies just among the orbitals that play an important role in generating the collectivity around $N=106$.

\begin{acknowledgments}  
Valuable discussions with K.~Matsuyanagi, M.~Matsuo, and Y.~R.~Shimizu are acknowledged. 
This work was supported by the JSPS KAKENHI (Grants No. 25287065, and No. 16K17687). 
The numerical calculations were performed on SR16000 and CRAY XC40 
at the Yukawa Institute for Theoretical Physics, Kyoto University, and 
on COMA (PACS-IX) at the Center for Computational Sciences, University of Tsukuba.
\end{acknowledgments}

\end{document}